\documentclass[pra,showpacs,twocolumn,superscriptaddress,longbibliography,amssymb]{revtex4-1}

\usepackage{graphicx}
\usepackage{bm}
\usepackage{amssymb}
\usepackage[normalem]{ulem}
\usepackage{soul, xcolor}
\setstcolor{blue}

\usepackage{dcolumn}

\usepackage{hyperref}
\hypersetup{
    colorlinks=true,
    linkcolor=blue,
    citecolor=blue,   
    urlcolor=blue
    }
    
\begin{document}

\title{Rotational magic conditions for ultracold molecules in the presence of Raman and Rayleigh scattering}

\author{Svetlana Kotochigova}
\email{skotoch@temple.edu}
\address{Department of Physics, Temple University, Philadelphia, Pennsylvania 19122, USA}
\author{Qingze Guan}
\address{Department of Physics, Temple University, Philadelphia, Pennsylvania 19122, USA}
\address{Department of Physics and Astronomy, Washington State University, Pullman, Washington 99164-2814, USA}
\author{Eite Tiesinga}
\address{Joint Quantum Institute, National Institute of Standards
and Technology and the University of Maryland, Gaithersburg MD 20899, USA}
\author{Vito Scarola}
\address{Department of Physics, Virginia Tech, Blacksburg, Virginia 24061, USA}
\author{Brian DeMarco}
\address{Department of Physics and IQUIST, University of Illinois at Urbana-Champaign, Urbana, IL 61801-3080, USA}
\author{Bryce Gadway}
\address{Department of Physics and IQUIST, University of Illinois at Urbana-Champaign, Urbana, IL 61801-3080, USA}
\date{\today}

\begin{abstract}
Molecules have vibrational, rotational, spin-orbit and hyperfine degrees of freedom or quantum states, each of which 
responds  in a unique fashion to external electromagnetic radiation. The  control over 
superpositions of these quantum states is key to coherent manipulation of molecules. For example, the 
better the coherence time the longer quantum simulations can last. The important quantity for 
controlling an ultracold molecule with laser light is its complex-valued molecular dynamic polarizability.  
Its real part determines the tweezer or trapping potential as felt by the molecule, while its imaginary part 
limits  the coherence time.
Here, our study shows that efficient trapping of a molecule in its vibrational ground state can be achieved by
selecting a laser frequency with a detuning on the order of tens of GHz relative to an
electric-dipole-forbidden molecular transition. Close proximity to this nearly forbidden transition allows to create a sufficiently deep  trapping potential for multiple rotational states without sacrificing
coherence times among these states from Raman and Rayleigh scattering. In fact, we demonstrate that magic trapping conditions for multiple rotational states 
of the ultracold $^{23}$Na$^{87}$Rb polar molecule can be created. 
\end{abstract}

\maketitle

\section{Introduction}

Electro-magnetic radiation plays a central role in the trapping and detection of ultracold molecules as well as the control of superpositions of their internal rovibrational and even external motional states. Some early highlights in trapping and state preparation for molecules can be found in 
Refs.~\cite{Kotochigova2006,Neyenhuis2012,Park2017,Frauke2018}. Optical tweezers and optical lattices, focussed and retero-reflected laser beams in the near infra-red or optical frequency domain, are the modern tools to trap molecules. For polar diatomic molecules, a natural choice for building a molecular quantum computer,  qubits are formed by pairs of rotational levels of their ground vibrational state.  Electric dipole-dipole interactions between such heteronuclear molecules lead to entanglement and are important for simulating many-body Hamiltonians. To make appropriate use of molecular rotation we must avoid dephasing due to light-induced rotational coupling. This may be achieved by choosing ``magic'' trapping conditions. In such traps, light-induced energy shifts of two or more rotational states are identical, eliminating dephasing associated with spatial variations in intensity across the trap.  Recently, this was achieved with  heteronuclear alkali-metal dimers \cite{Kotochigova2010, Qingze2021, He2021}.

The guiding principles of selecting trapping laser frequencies for ultracold ground-state molecules are to chose the frequency either well below the minimum energy of the electronically excited potentials or in the near-resonant region of mostly forbidden molecular transitions. The first idea was realized in multiple experiments using YAG lasers with wavelengths around 1064 nm \cite{Science08, Kotochigova2010, Neyenhuis2012}. These studies with off-resonant light have also shown that magic conditions can only exist  as long as the angles between the laser polarization and either the direction of an external static magnetic or electric field  are carefully controlled. In fact, the authors of Ref.~\cite{Neyenhuis2012} showed in measurements of the AC polarizability and of the coherence time of microwave transitions between rotational states that there exists an optimal angle between the orientation of the light polarization and a magnetic field.

The second approach for the  construction of rotational magic traps  was pioneered for $^{23}$Na$^{40}$K~\cite{Bause2020}, $^{87}$Rb$^{133}$Cs~\cite{Qingze2021,Gregory2024}, and $^{23}$Na$^{87}$Rb~\cite{Lin2021,He2021}. In all cases, the authors used a laser with a carefully chosen frequency between two vibrational levels  of the excited b$^3\Pi_{0^+}$ molecular electronic state. 
Electric dipole transitions from the ground X$^1\Sigma^+$ state to this electronic state are narrow and almost forbidden.
The laser detuning from vibrational levels is relatively small on the order of tens of GHz. Most importantly, external electric or magnetic fields are not required. Nevertheless, this second approach can be readily applied to related diatomic and  polyatomic molecules. 

This paper describes quantitative theoretical calculations aimed at finding ``magic'' conditions 
 for ultracold $^{23}$Na$^{87}$Rb molecules prepared in superpositions of rotational states of its vibrational ground state and held in place by tweezer forces. We focus on the second approach using a frequency region just above the minimum of the potential of the exited b $^3\Pi_{0^+}$ molecular state in the presence of a 335~G magnetic field defining a natural quantization direction. This magnetic field strength, corresponding to the field location of a Fano-Feshbach resonance in the collision of ultracold $^{23}$Na and $^{87}$Rb atoms, was inspired by the experiments described in Ref.~\cite{Wang2016a}. 
No electric field is present and the polarization of the tweezer light is linear and along the magnetic field direction contrasting with the conditions studied in Ref.~\cite{He2021}. We have also developed an approach to enhance coherence times for superpositions of more than two rotational states and derive approximate analytical expressions for the dynamic polarizabilities in order to better understand the origin of  magic conditions. These expressions will enable experimentalists to analyze magic conditions for their own molecular system and experimental conditions. 

Unavoidable scatter of photons out of laser beams by a molecule leads to to detrimental off-resonant scattering and decoherence. The relevant processes are then classified as either Raman or Raleigh scattering. In both processes scattering corresponds to the off-resonant absorption of a laser photon by the molecule, promoting the molecule to an electronically excited state, followed by spontaneous emission of a photon into a bath mode. In this paper, we determine rates for molecular Raman and Rayleigh photon-scattering processes corresponding to the stimulated absorption of a laser photon by the molecule, promoting the molecule to an electronically excited state, followed by spontaneous emission of a photon.  For Raman scattering the initial and final rovibrational molecular states are different whereas for Rayleigh scattering the initial and final state are the same. These processes define the ultimate limit on coherence times. For laser-cooled atoms, these scattering processes  were studied in Refs.~\cite{Cline1994,UysPRL2010,Brown2018}. In this paper, we  give for the first time guidance  to experimentalists on how to minimize all or parts of these decoherence processes for heteronuclear molecules. 

\section{Magic trapping frequencies near  $b^3\Pi_{0^+}$, $v'$ = 0 rotational resonances}

We begin by calculating the dynamic polarizability or AC Stark shift $\alpha_{{\rm X},vJM}(\nu,\vec{\varepsilon})$ for rotational levels  ${J=0}$ to $5$ of the ${v=0}$ vibrational level $|{\rm X}, vJ M\rangle$ of the ground  X$^1\Sigma^+$ state of the $^{23}$Na$^{87}$Rb molecule  in the absence of external electric or magnetic fields and without molecular spin-rotation, hyperfine, and Zeeman interactions. Here, $M$ is the projection quantum number of angular momentum $J$ along a laboratory or space-fixed axis to be defined later. The tweezer laser with laser frequency $\nu$ is linearly polarized along space-fixed direction $\vec\varepsilon$ throughout this paper.  We then determine  laser light frequencies that allow  simultaneous magic trapping of multiple rotational states using light nearly resonant with rovibrational levels of the b$^3\Pi_{0^+}$ electronic state. Transitions between the X$^1\Sigma^+_{0^+}$ and b$^3\Pi_{0^+}$ states are weak and only allowed through weak spin-orbit coupling with the A$^1\Sigma^+_{0^+}$ state. 
See Ref.~\cite{Herzberg} for a description of molecular notation.

\begin{figure}
\includegraphics[scale=0.28, trim=10 15 0 0,clip]{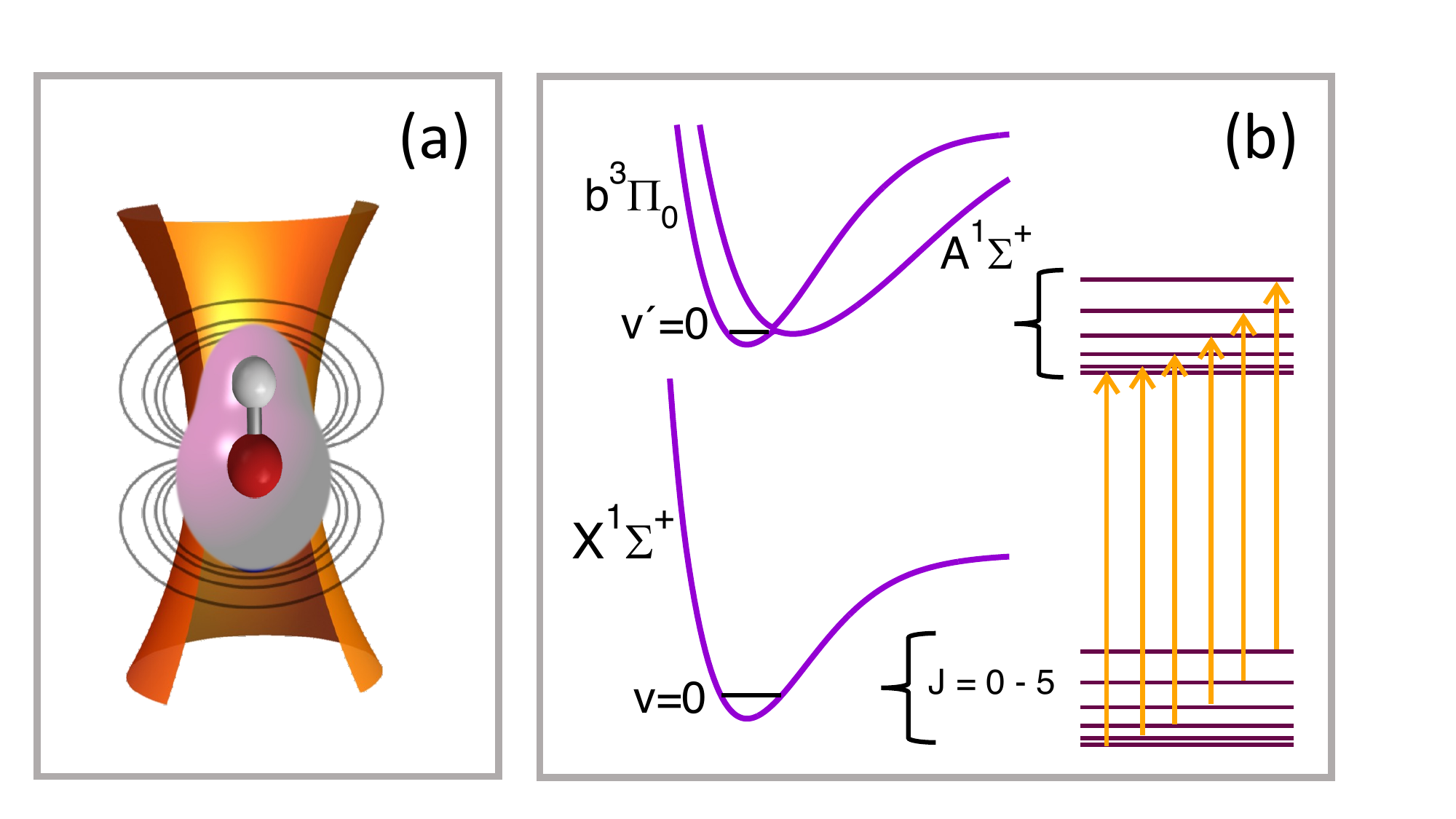} 
\caption{(a) Schematic presentation of  ground-state heteronuclear NaRb  trapped in an optical tweezer potential. (b)  The potential energies (purple curves) as function of atomic separation of the three most-important for us electronic states of NaRb. The two black horizontal lines in the potentials represent the energetically lowest $v=0$ and $v'=0$ vibrational levels of the X$^1\Sigma^+$ state and the coupled A$^1\Sigma^+$ and b$^3\Pi_{0^+}$ complex, respectively. Relevant rotational levels $J=0$ to 5 for both of these vibrational states are shown on the right. Near resonant optical transitions, orange lines with arrows, are used in a search for magic conditions as a function of  tweezer laser frequency.}
\label{scheme}
\end{figure}

Figure~\ref{scheme}(a) schematically shows  the NaRb molecule trapped in a tweezer potential, while Fig.~\ref{scheme}(b) displays the three relevant relativistic Hund's case (c) $\Omega^\sigma=0^+$ potential energy curves of the NaRb molecule, where $\Omega$ is the absolute value of the projection quantum number of the total electronic angular momentum along the internuclear axis and ${\sigma=\pm}$ represents the symmetry of  ${\Omega=0}$ electronic wavefunctions under reflection through a plane containing the internuclear axis. More precisely, the  excited non-relativistic Hund's case (a) A$^1\Sigma^+_{0^+}$ and b$^3\Pi_{0^+}$ states are coupled by the spin-orbit interaction, which leads to two relativistic $\Omega=0^+$ adiabatic potentials that have a narrow avoided crossing near interatomic separation $R\approx R_{\rm c}=7.5a_0$~\cite{Docenko2007}, where $a_0=0.0529$ nm is the Bohr radius. The energetically lowest $\Omega=0^+$ rovibrational states near the bottom or minimum of the nominally b$^3\Pi_{0^+}$ potential  have  a small admixture of the A$^1\Sigma^+_{0^+}$ state.
As  electric dipole transitions between the singlet X$^1\Sigma^+$ and triplet b$^3\Pi_{0^+}$ states are forbidden, this leads to weak, 
but easily observable transitions from rovibrational levels of the X$^1\Sigma^+$, $\Omega=0^+$  state.  We observe that the equilibrium separations and harmonic frequencies of the X$^1\Sigma^+$ and b$^3\Pi_{0^+}$ states are almost the same. We have used the non-relativistic X$^1\Sigma^+$,  A$^1\Sigma^+_{0^+}$ and b$^3\Pi_{0^+}$ potentials, spin-orbit matrix elements, and the X$^1\Sigma^+$ to  A$^1\Sigma^+$ electronic transition dipole moment  as functions of $R$ given by Refs.~\cite{Pashov2005, Docenko2007}. 

For rotational states $J,M$ of the ${v=0}$ X$^1\Sigma^+$ ground-state of NaRb, the calculation of the sum over intermediate, excited states that appears in the evaluation of $\alpha_{{\rm X},vJM}(\nu,\vec{\varepsilon})$ can be simplified. 
The relevant laser frequencies are {\it nearly resonant} with rovibrational levels near the minimum of the b$^3\Pi_{0^+}$  potential. Consequently, we separate  $\alpha_{{\rm X},vJM}(\nu,\vec{\varepsilon})$ into two contributions. As a first contribution we will use the  corresponding vibrationally averaged transition dipole moments to construct this contribution to  $\alpha_{{\rm X},vJM}(\nu,\vec{\varepsilon})$. A  second contribution to $\alpha_{{\rm X},vJM}(\nu,\vec{\varepsilon})$  is due to {\it off-resonant} transitions to other electronic states. They are computed within a quasi-static approximation, 
where  so-called parallel and perpendicular polarizabilities $\alpha_{\parallel}(\nu,R)$ and $\alpha_{\perp}(\nu,R)$, corresponding to laser polarizations parallel and perpendicular  to the body-fixed internuclear axis, respectively,
are computed  as functions of laser frequency $\nu$ and atom-atom separation $R$ near the equilibrium separation $R_{\rm e}$ of the X$^1\Sigma^+$ state using  the linear response theory formulation within  software package Q-Chem \cite{Q-Chem}. Q-Chem computes electronic states based on a non-relativistic description of the electrons.
In practice, we realize that these two quasi-static polarizabilities are to good approximation independent of $R$
over the radial width of the ${v=0}$ vibrational level of the X$^1\Sigma^+$ state. We thus only  compute
$\alpha_{\parallel}(\nu,R)$ and $\alpha_{\perp}(\nu,R)$ at $R=R_{\rm e}$ and drop argument $R$ for the remainder of this article.

The two quasi-static polarizabilities of the X$^1\Sigma^+$ state  have been obtained with a non-relativistic configuration-interaction electronic-structure calculation using an all-electron basis set for Na. Single and double excitations were allowed from these basis functions. An effective core potential describes the 28 inner electrons of Rb. Single and double  excitations were allowed for the remaining electrons of Rb.

\begin{figure}
\includegraphics[scale=0.38, trim=0 0 0 0,clip]{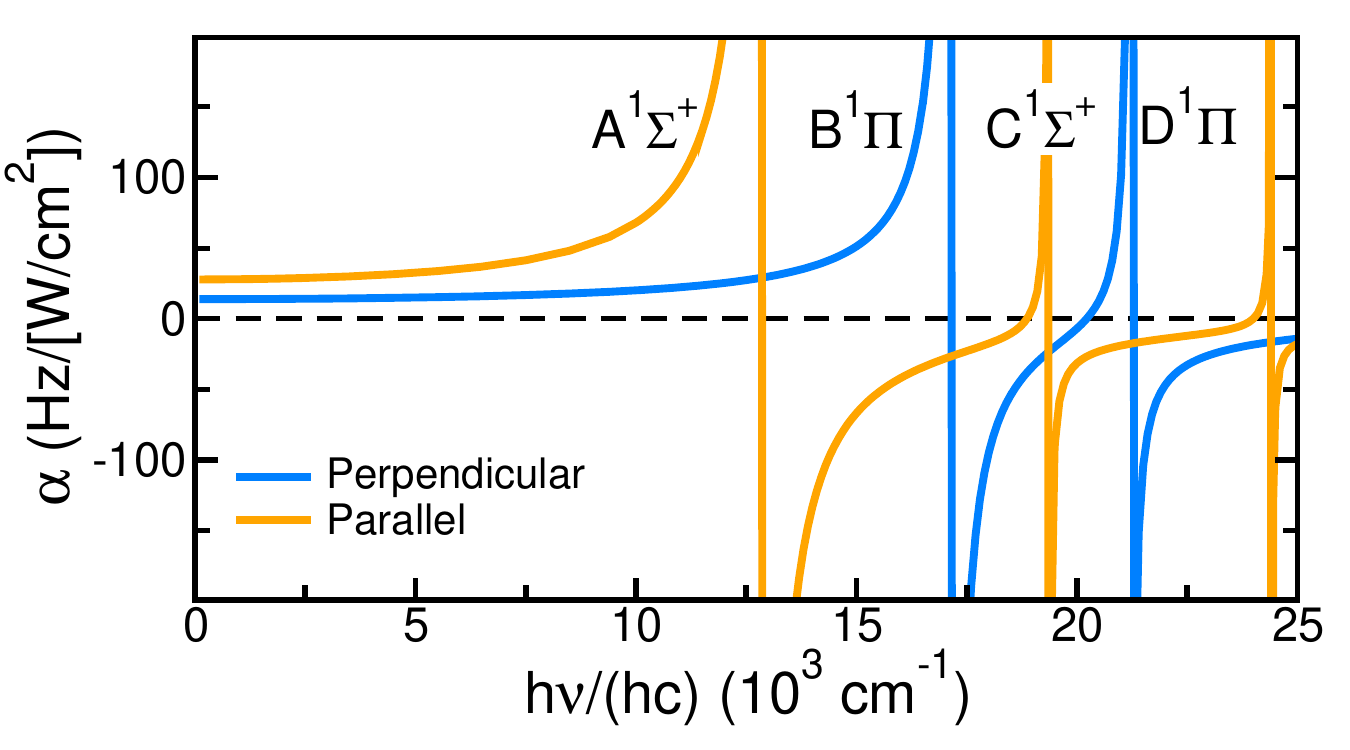} 
\caption{The quasi-static parallel (orange curve) and perpendicular (blue curve) electronic polarizabilities of the X$^1\Sigma^+$ state  at its equilibrium separation $R_{\rm e}=6.885a_0$  and  photon energies up to $hc\times 25\,000$ cm$^{-1}$. The energetically lowest four resonances are labeled with state $^1\Lambda$.
The data is based on non-relativistic configuration-interaction calculations with the Q-Chem software package.}
\label{polariz}
\end{figure}

Figure~\ref{polariz} shows the two quasi-static polarizabilities of the X$^1\Sigma^+$ state of NaRb at 
${R=R_{\rm e}}$ as functions  of photon energy from zero to $hc\times25\,000$ cm$^{-1}$.  Here, $h$ is the Planck constant and $c$ is the speed of light in vacuum. Over this large photon-energy range, several resonances are visible. Each  corresponds to a  transition between the X$^1\Sigma^+$ state and a $^1\Lambda$ state.  In fact, in our non-relativistic  formulation,  $\alpha_{\parallel}(\nu)$ only contains contributions from transitions  to  singlet $^1\Sigma^+$ electronic states,  while $ \alpha_{\perp}(\nu)$  only contains contributions from transitions  to singlet $^1\Pi$ states.  Finally,  the quasi-static contributions to the polarizability of levels of the X$^1\Sigma^+$ state  have  a small photon-energy dependence for photon energies near the minimum of the triplet b$^3\Pi_{0^+}$  potential.

The resonant contribution to the polarizability has been determined in several steps.
We compute two-channel radial eigenvalues and eigenfunctions of the spin-orbit coupled and shifted $\Omega=0^+$ A$^1\Sigma^+$ and b$^3\Pi_{0^+}$  states for total angular momentum $J'=0$, 1, \dots, 6 
using a discrete variable representation (DVR) of the radial relative kinetic energy operator \cite{Colbert1992}.  For each $J'$, the eigenvalues  $E_{{\rm Ab},v'J'}$ are labeled $v'=0$, 1, $\dots$ with increasing energy and wavefunctions of the energetically lowest  $v'$ levels are to good approximation b$^3\Pi_{0^+}$ states. The eigenenergies are independent of projection quantum number $M'$ of $J'$. 

We  then compute rovibrational wavefunctions $v$ and energies $E_{{\rm X},vJ}$ of the X$^1\Sigma^+$ state for  $J$ up to 5 with the same DVR and radial grid used to compute eigenpairs of the coupled A$^1\Sigma^+$ and b$^3\Pi_{0^+}$ system. The eigenenergies are independent of projection quantum number $M$ of $J$. The use of the same radial grid avoids interpolation of wavefunctions in the  computation of  vibrationally-averaged transition dipole moments
using the $R$-dependent transition dipole moment between the X$^1\Sigma^+$ and A$^1\Sigma^+$ states.

Finally, we compute the resonant-part of the polarizability $\alpha_{{\rm X},vJM}(\nu,\vec{\varepsilon})$ of $v=0, JM$ X$^1\Sigma^+$  states using only the energetically lowest rovibrational levels of the coupled A$^1\Sigma^+$-b$^3\Pi_{0^+}$ system that have a large b$^3\Pi_{0^+}$ character and thus a small vibrationally-averaged transition dipole moment. The choice to limit the determination of the resonant-part of the polarizability to a few levels of the  A$^1\Sigma^+$-b$^3\Pi_{0^+}$ system avoids double counting the effects of the A$^1\Sigma^+$ state
when combining the resonant and off resonant contributions of $\alpha_{{\rm X},vJM}(\nu,\vec{\varepsilon})$.
In principle, the projection degeneracy is broken by hyperfine interactions between  nuclear quadrupole moments and the rotation of the molecule as well as Zeeman interactions for the nuclear spins. The nuclear spin of both $^{23}$Na and $^{87}$Rb is 3/2. However, the hyperfine splittings for the $\Omega=0^+$ states, as shown in Ref.~\cite{Wang2016a},
are small  compared to the rotational energies described below. Here, we omit the effects of  hyperfine interactions on magic conditions. 

\begin{figure*}
\includegraphics[scale=0.5,trim=10 100 0 0,clip]{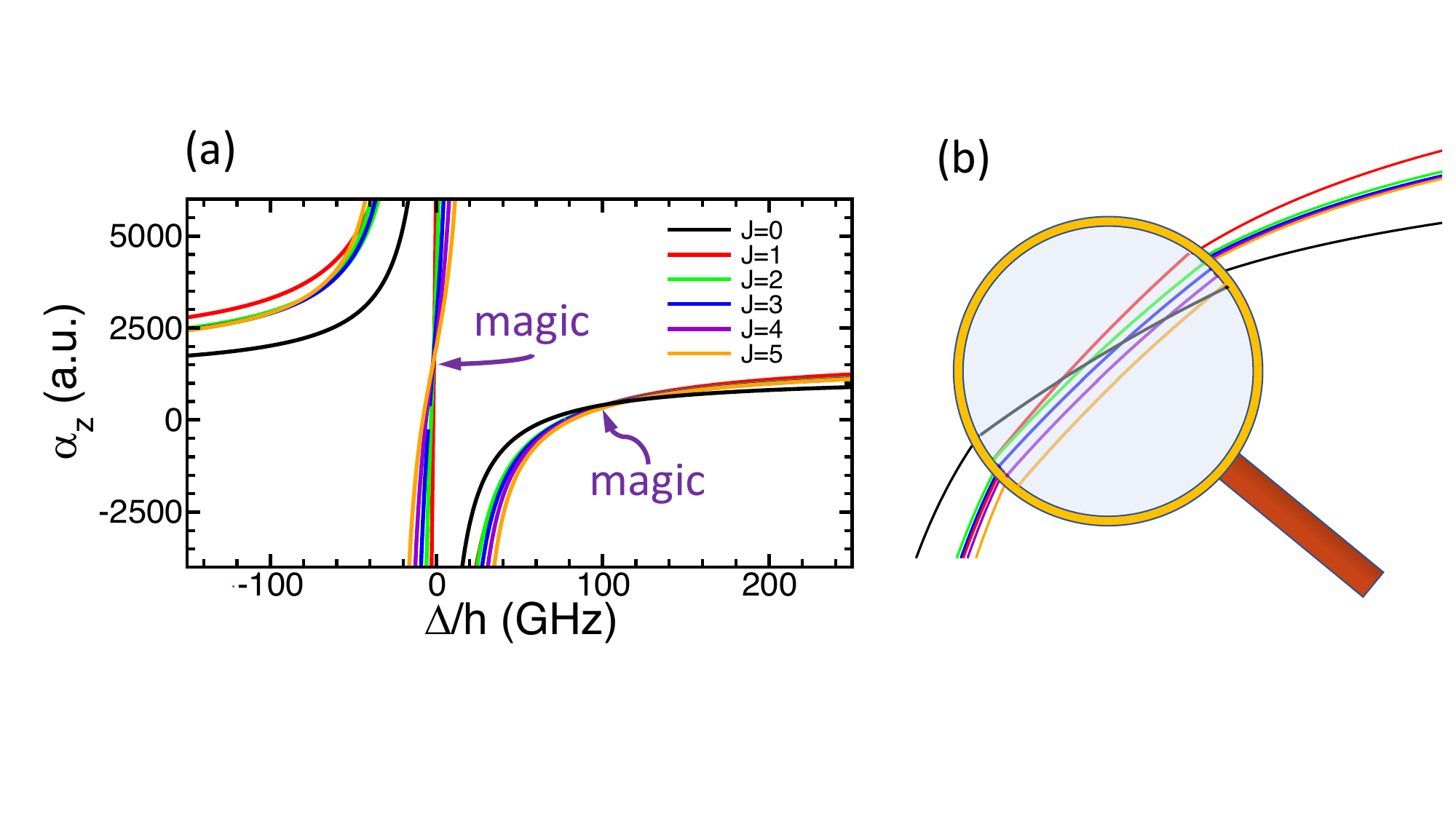}
\caption{(a) $^{23}$Na$^{87}$Rb dynamic  polarizabilities of rotational  $v = 0$ levels of the X$^1\Sigma^+$ state in atomic units for $z$-linear polarized light  as functions of laser frequency detuning $\Delta/h$ near transitions from the $v = 0$  X$^1\Sigma^+$ state to the $v' = 0$  level of the coupled A$^1\Sigma^+$-b$^3\Pi_{0^+}$ system. Each curve corresponds to the dynamic  polarizability of a different rotational level $J$ with $M=0$.
Zero detuning  $\Delta$ corresponds to the resonant transition from $v = 0$, $J=0$ level of the X$^1\Sigma^+$ state to the $v' = 0, J'=1$ level of the coupled A$^1\Sigma^+$-b$^3\Pi_{0^+}$ system. The purple arrows indicate magic detunings, where  rotational states have the same or nearly the same polarizabilities. (b) Schematically magnified region of  dynamic polarizabilities, near the frequency detuning $\Delta/h=100$ GHz. The colors of the curves are the same as those in panel (a). We observe that the curve for $J=0$ crosses those for $J>0$.  }
\label{magic}
\end{figure*}

Figure~\ref{magic}  shows  dynamic polarizabilities of the  $v=0$, $J=0$, 1,\dots, 5  rotational levels of the $^{23}$Na$^{87}$Rb X$^1\Sigma^+$ state with projection quantum number $M=0$ in a 335 Gauss magnetic field parallel  to the polarization of linearly polarized light as functions of laser frequency in the neighborhood of the $v' = 0$ level of the coupled A$^1\Sigma^+$-b$^3\Pi_{0^+}$ system. There is no  external electric field applied. 
The dynamic polarizabilities include both  resonant and off-resonant contributions and
we chose the quantization axis of the molecular angular momentum $J$ and $J'$ in the same direction as that of the laser polarization. 
The horizontal axis gives photon energy detuning $\Delta= h\nu-{\cal E}_{v'=0}$, where molecular transition energy ${\cal E}_{v'=0}=E_{{\rm Ab},v'=0,J'=1}-E_{{\rm X},v=0,J=0}$. 
Our estimate provides that ${\cal E}_{v'=0} = hc\times11\,306.1$ cm$^{-1}$, which corresponds to a laser wavelength close to $884.6$~nm. These results are in a good agreement with the experimental measurements and theoretical calculations of Ref.~\cite{He2021}.

The curves for $J=0$ and $J>0$ in Fig.~\ref{magic} have different behaviors. The polarizability for the $J=0$
has a single resonance at $\Delta=0$. Those for $J>0$ have two resonances
located at $L_{J}(v=0,v'=0)$ and $R_{J}(v=0,v'=0)$, respectively,
where
\begin{eqnarray}
L_{J}(v,v')= J(J+1) B_{v} - [J(J-1)-2]B_{v'}
\end{eqnarray}
and
\begin{eqnarray}
R_{J}(v,v')=J(J+1) B_{v} - [(J+1)(J+2)-2]B_{v'}
\end{eqnarray}
with rotational constants $B_{v}$ and $B_{v'}$ for the $v$ vibrational level of the X$^1\Sigma^+$ state and the $v'$ vibrational level of the coupled A$^1\Sigma^+$-b$^3\Pi_{0^+}$ system, respectively. For $^{23}$Na$^{87}$Rb, $B_{v=0}/hc=0.069\,70$ cm$^{-1}$ and $B_{v'=0}/hc=0.069\,88$ cm$^{-1}$. The two values agree to better than 0.5\,\%. These equations follow from photon selection rules $|J-1|\le J'\le J+1$ and $|J'-J|$ is odd.

Panel (a) of Fig.~\ref{magic} shows that there exist two {\it magic} laser frequencies. The first is located near 
$\Delta/h = -2$ GHz and $J>0$  rotational levels have nearly the same polarizabilities. The polarizability for $J=0$ is large and positive at this detuning. The second is located
near $\Delta/h=+100$ GHz, all $J=0, \dots,5$ rotational levels have nearly the same polarizabilities. Panel (b) looks in more detail at the latter frequency region. In particular,  the polarizabillity of the $J=0$ level is equal to that of the $J=1$, 2, 3, 4, 5 rotational levels at detuning $\Delta/h = 103$ GHz, $105$ GHz, $108$ GHz, $112$ GHz, and $116$ GHz, respectively. In fact, the differences between these $\Delta$ with $J>0$  is increasing with $J$. The origin of this effect lies in the increase of rotational energies with $J$.

\section{Analytical results for magic polarizability} 

We  find it useful to derive analytical expression for the dynamic polarizabilities  shown in Fig.~\ref{magic} in order to better understand the origin of magic conditions as well as simplifying their determination for wide variety of molecules. 
 For  rovibrational level  $v=0$ $JM$ of the X$^1\Sigma^+$ state, the dynamic polarizability for linearly-polarized laser light near  transitions to   vibrational states $v'=0$ or $v'=1$ of the coupled A$^1\Sigma^+$-b$^3\Pi_{0^+}$ system is  well approximated by
\begin{eqnarray}
\label{general}
\lefteqn{\alpha_{{\rm X},v=0JM}(\nu,\vec\varepsilon) = - \frac{3\pi c^2}{2\omega_{v'}^3}
\left[A_{J,M}(\theta_{\rm p}) \frac{\hbar\Gamma_{0,v'}}{\Delta_{v'} +  L_{J}(0,v')}  \right.} \\
&&\qquad\qquad\qquad \qquad \qquad  \left.+ B_{J,M}(\theta_{\rm p}) \frac{\hbar\Gamma_{0,v'}}{\Delta_{v'} + R_{J}(0,v')}\right]
\nonumber \\
&& 
+ \left[A_{J,M}(\theta_{\rm p}) + B_{J,M}(\theta_{\rm p})\right](\alpha_{\text{bg}, \parallel}-\alpha_{\text{bg}, \perp})
+ \alpha_{\text{bg}, \perp}\,, \nonumber
\end{eqnarray}
where $\theta_{\rm p}$ is the angle between the polarization of the laser $\vec\varepsilon$ with respect to the
quantization axis for the molecular states, our laboratory-fixed $z$ axis. 
The energy detuning 
\begin{equation}
\Delta_{v'}= h\nu-\hbar\omega_{v'}
\end{equation}
with $\hbar\omega_{v'}=E_{{\rm Ab},v',J'=1}-E_{{\rm X},v=0,J=0}$. The  terms on the first two lines of  Eq.~\ref{general} lead to resonances in the dynamic polarizability. In fact,
there are one and two resonances for $J=0$ and $J>0$, respectively.
The $\Gamma_{0,v'}$ are linewidths of the vibrational levels $v'$  of the coupled A$^1\Sigma^+$-b$^3\Pi_{0^+}$ system.

The parallel and perpendicular $\alpha_{\text{bg},\parallel}$ and $\alpha_{\text{bg},\perp}$ are body-fixed
background polarizabilities.
The dimensionless angular factor $A_{J,M}(\theta_{\rm p})$ is given by
\begin{eqnarray}
\label{factor_A}
A_{J,M}(\theta_{\rm p}) &=&
\frac{J(J+1)-3M^2}{2(2J+1)(2J-1)} \cos^2{\theta_{\rm p}} \\
  && \qquad + \frac{(J-1)J+M^2}{2(2J+1)(2J-1)} \nonumber
\end{eqnarray}
for $ |M| < J$ and
\begin{equation}
A_{J,M}(\theta_{\rm p}) =\frac{(J+|M|)(J+|M|-1)}{4(2J+1)(2J-1)}\sin^2{\theta_{\rm p}}
\end{equation}
for $|M| = J$. Note that $A_{J,M}(\theta_{\rm p})=0$ for $J=0$. 
Finally, the dimensionless   $B_{J,M}(\theta_{\rm p})$ is given by
\begin{eqnarray}
\label{factor_B}
B_{J,M}(\theta_{\rm p}) &=& \frac{J(J+1)-3M^2}{2(2J+1)(2J+3)}\cos^2{\theta_{\rm p}} \\
  && \quad +\frac{(J+1)(J+2) + M^2}{2(2J+3)(2J+1)}\,.\nonumber
\end{eqnarray}

Equation \ref{general} can only be used for energy detunings that are much smaller than the vibrational spacing between different $v'$ states of the A$^1\Sigma^+$-b$^3\Pi_{0^+}$ system.
On the other hand, the energy detunings must be much larger than any hyperfine  and Zeeman splittings
in the coupled A$^1\Sigma^+$-b$^3\Pi_{0^+}$ system and for energy detunings much larger than $\hbar\Gamma_{0,v'}$.

Finally, a Taylor expansion of the right hand side of Eq.~\ref{general}  assuming $|\Delta_{v'}|\gg |L_J|$ and $|\Delta_{v'}|\gg |R_J|$ gives
\begin{eqnarray}
\label{general_app}
\lefteqn{\alpha_{{\rm X},v=0JM}(\nu,\vec\varepsilon) =
\left[A_{J,M}(\theta_{\rm p}) + B_{J,M}(\theta_{\rm p})\right]}\\
&& \times
\left(-\frac{3\pi c^2}{2\omega_{v'}^3}\frac{\hbar\Gamma_{0,v'}}{\Delta_{v'}}
+\alpha_{\text{bg}, \parallel} - \alpha_{\text{bg}, \perp}\right)+\alpha_{\text{bg}, \perp}, 
+\cdots. \nonumber
\end{eqnarray}
From an inspection of Eq.~\ref{general_app}, we  realize that we can always find an energy detuning independent of $\theta_{\rm p}$ and  $J$ such that the term in parenthesis  vanishes. At this energy detuning the dynamic polarizability is $\alpha_{\text{bg},\perp}$, the same for all $\theta_{\rm p}$ and  $J$ within our approximations, and the optical trap is magic for all rotational states.  Higher-order terms in Eq.~\ref{general_app} will add small $\theta_{\rm p}$- and  $J$-dependent corrections and are observed in Fig.~\ref{magic}.

\section{Vibrationally-resolved imaginary polarizabilities of the X$^1\Sigma^+$ state} 

In addition, we developed an approach based on ideas of Refs.~\cite{Kotochigova2006} to evaluate the  imaginary dynamic polarizabilities of rovibrational levels of the ground state in vicinity of rovibrational levels of the excited state potentials. The imaginary part of $\alpha_i$ describes incoherent decay that leads to loss of molecules from the optical trap. Our calculation is based on perturbation theory method with specific focus on relativistic spin-orbit coupling between A$^1\Sigma^+$-b$^3\Pi_0$ complex. The simulations are performed using electronic potentials, permanent, and transition dipole moments of NaRb determined in Refs.~\cite{Pashov2005, Docenko2007, Dulieu2017}. 

We analyze the imaginary dynamic polarizability in the wide range laser frequency from 10000 to 20000 cm$^{-1}$ that is in the resonance with multiple vibrational levels of the excited state potentials.

The  molecular dynamic polarizability $\alpha_i(h\nu,\vec{\epsilon})$ at  frequency $\nu$ and laser polarization $\vec{\epsilon}$ of state $i$ is complex valued. To good approximation its imaginary part is
\begin{eqnarray}
\label{imagin}
 {\rm Im}\left[ \alpha(h\nu,\vec{\epsilon}) \right]&=&
 - \frac{1}{\epsilon_0c}
   \sum_{f} \frac{\hbar\gamma_f/2 }{(E_f  - E_i)^2 - (h\nu)^2}\\
    && \qquad \qquad \times  |\langle f| d(R) \hat{R}\cdot \vec{\epsilon}\,|i\rangle|^2\,, \nonumber
\end{eqnarray}
where kets $|i\rangle $ and $|f\rangle$ are simplified labels for  initial 
rovibrational wavefunctions of the X$^1\Sigma^+$ potential  and those of excited electronic states, respectively.
Their energies are $E_i$ and $E_f$, respectively, and  $\gamma_f$ is the natural line width of excited rovibrational levels.

\begin{figure}
\includegraphics[scale=0.32,trim=0 20 0 10,clip]{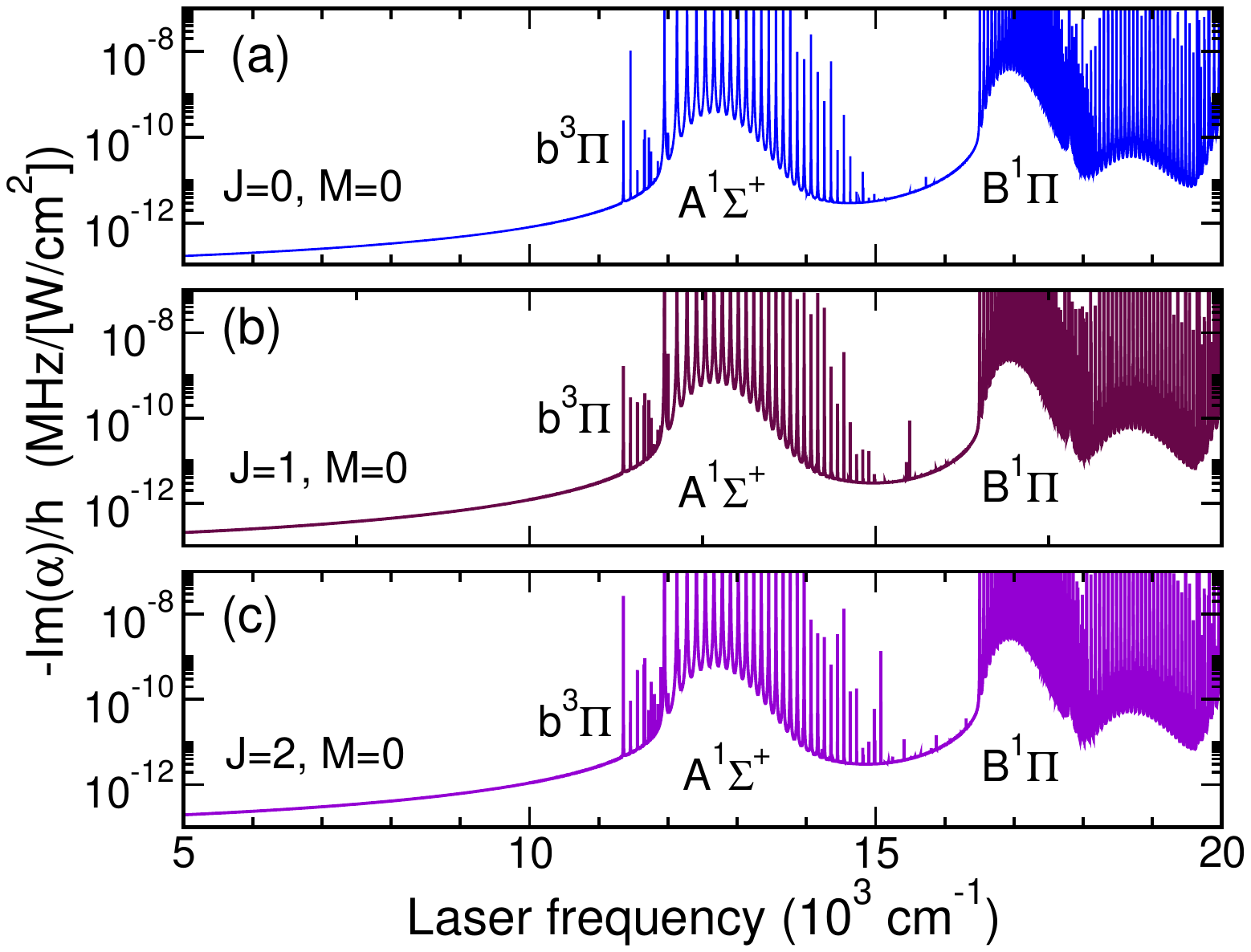}
\vspace{-6mm}
\caption{Minus one times the imaginary part of the dynamic polarizabilities of the $J=0$, $1$, and 2 rotational levels of the vibrational ground state of 
        $^{23}$Na$^{87}$Rb with projection quantum number  $M=0$ along the $z$ axis as functions of laser frequency  in panels (a), (b), and (c), respectively. Imaginary polarizabilities are presented for laser polarization $\sigma_s$ along the $\hat z$ direction.}
\label{ImaginPolar}
\end{figure}

Figure~\ref{ImaginPolar}, panels (a), (b), and (c) demonstrate imaginary dynamic polarizabilities of the $J=0, 1$, and $2$ rotational levels, respectively, of X$^1\Sigma^+$ with projection quantum number $M=0$. It turns out that, to good approximation, the three curves are the same except for a frequency-independent scale factor. Deviations from these scalings occur very close to the resonances, i.e. on the order of the rotational spacing of the molecule. We calculate imaginary polarizabilities for these levels taking into account the rovibrational structure of lower excited states in the units of MHz/[W/cm$^2$], which are often used in experimental measurements. Note that one atomic unit of polarizability corresponds to 4.68645 $\times$ 10$^{-8}$ MHz/[W/cm$^2$].

We evaluated the rovibrational molecular line widths of excited electronic states dissociating to either a singly-excited Na or Rb atom by first computing a $R$-dependent optical potential $-i\Gamma(R)/2$ \cite{Zygelman1988} for each excited electronic state. Here, $\Gamma(R)$ is positive and proportional to $|\delta E(R)|^3d^2(R)$, where $\delta E(R)$ and $d(R)$ at internuclear separation $R$ are the potential energy difference  and the  transition electronic dipole moment between an excited state and the ground state, respectively. Finally, the  energies $E_i$ and $E_f$ and  line widths $\gamma_f$ of  rovibrational levels of   electronic states were found by computing radial  rovibrational wavefunctions, energies, and  matrix elements of $\Gamma(R)$. By construction, the  imaginary part  is  negative. It is seven orders of magnitude smaller than the real part. For $J=1$ and $2$, $M=0$ the polarizability  depends  on the polarization direction of the trapping light. 

The imaginary part of the polarizabilities are slowly varying with frequency in regions outside multiple closely spaced resonant  features, where $\alpha(h\nu,\vec{\epsilon})$ is orders of magnitude larger than in the slowly varying regions. The resonant like features are due to the rovibrational bound states of excited electronic potentials. In fact, we could assign the resonances as due to the b$^3\Pi$, A$^1\Sigma^+$, and B$^1\Pi$ states. These resonances are strongest when the inner- or outer-turning point of rovibrational wavefunctions of the excited electronic potentials coincides with the equilibrium separation of the X$^1\Sigma^+$ potential.

The calculations of the imaginary part of the polarizability allowed us predict the role of unwanted decoherence processes.  In particular, optical fields  can transfer population from a rovibrational level of the electronic ground state  to  rovibrational levels of an excited electronic state, which then by the spontaneous emission  decays to many rovibrational levels of the X$^1\Sigma^+$ state. As a result, we lose control over the molecule.

\section{Effect of Raman and Rayleigh scattering on the magic conditions} 

Spontaneous photon scatter after the absorption of an optical laser photon leads to 
loss of information about the ro-vibrational state of the X$^1\Sigma^+$ molecule as scattered photons remain undetected. 
Two processes can occur. For Raman scattering the initial $vJM$
 and  final  $v'J'M'$ ro-vibrational levels of the X$^1\Sigma^+$ molecular state have a different energy. For Raleigh scattering the initial and final
state have the same energy and the spontaneously-emitted photon only differs
in direction from that of the absorbed laser photon. 

The rate for Raman and Rayleigh processes is proportional to the
square of a sum of two-photon transition amplitudes and inteferences play a prominant role.
\cite{UysPRL2010,Cline1994}. 
Following the derivations in Refs. \cite{Loudon, UysPRL2010,Cline1994} in SI units the state-to-state {\it off-resonant} decay rate is
\begin{eqnarray}
  \gamma_{{\rm X},v'J'M'\leftarrow {\rm X},vJM} &=& n_\nu\, c\,\sigma_{{\rm X},v'J'M'\leftarrow {\rm X},vJM}  \nonumber\\
     &=& \frac{ \sigma_{{\rm X},v'J'M'\leftarrow {\rm X},vJM}}{h\nu}\ I
\end{eqnarray}
with laser-photon number density $n_\nu$ of photons with energy $h\nu$ and polarization $\epsilon$,  and laser intensity $I =h\nu n_\nu c$. Moreover, the state-to-state resolved cross section 
 \begin{equation}
\sigma_{{\rm X},v'J'M'\leftarrow {\rm X},vJM}= \frac{8\pi}{3}  
\alpha^4 a_0^2   \frac{h\nu (h\nu')^3}{E_{\rm h}^4} |{\cal G}|^2\,,
  \end{equation}
where $\alpha$ is the fine structure constant, $E_{\rm h}$ is the Hartree energy,  
and dimensionless
\begin{widetext} 
 \begin{eqnarray}
 |{\cal G}|^2&=&\frac{E_{\rm h}}{(ea_0)^2}
 \sum_{q=-1}^1  \left|\sum_{e, v'' J'' M''}\left[ \frac{ \langle {\rm X}, v'J'M' | d_q | e, v''J''M''\rangle
  \langle e, v''J''M'' | \vec d \cdot \vec \epsilon \,| {\rm X}, vJM \rangle}{E_{{\rm X},vJM}+h\nu-E_{e,v''J''M''}} \right. \right.\nonumber\\
    &&  \qquad\qquad\qquad\qquad\qquad\qquad
    \left. \left. + \frac{ \langle {\rm X}, v'J'M' | \vec d \cdot \vec\epsilon\, | e, v''J''M''\rangle
  \langle e, v''J''M'' | d_q | {\rm X}, vJM \rangle
  }{E_{{\rm X},vJM}-E_{e,v''J''M''}-h\nu'} \right]
    \right|^2 \,,
 \end{eqnarray} 
 \end{widetext}
where $h\nu'=E_{e,v''J''}-E_{{\rm X},v'J'}$  is the energy of the spontaneously emitted photon and
$E_{{\rm X},v'J'M'}-E_{{\rm X},vJM}\le h\nu$.
The sum inside the absolute value is over all rovibrational levels
and continuum states $v''J''M''$ of electronically excited states $e$.
In practice, for pair $M$ and $M'$ and $ \vec\epsilon=\hat z$ only one projection $q$ contributes in the sum over $q$
and the second term in the square brackets is negligible for laser frequencies in the optical domain.

\begin{figure}
\includegraphics[scale=0.27,trim=0 0 0 0,clip]{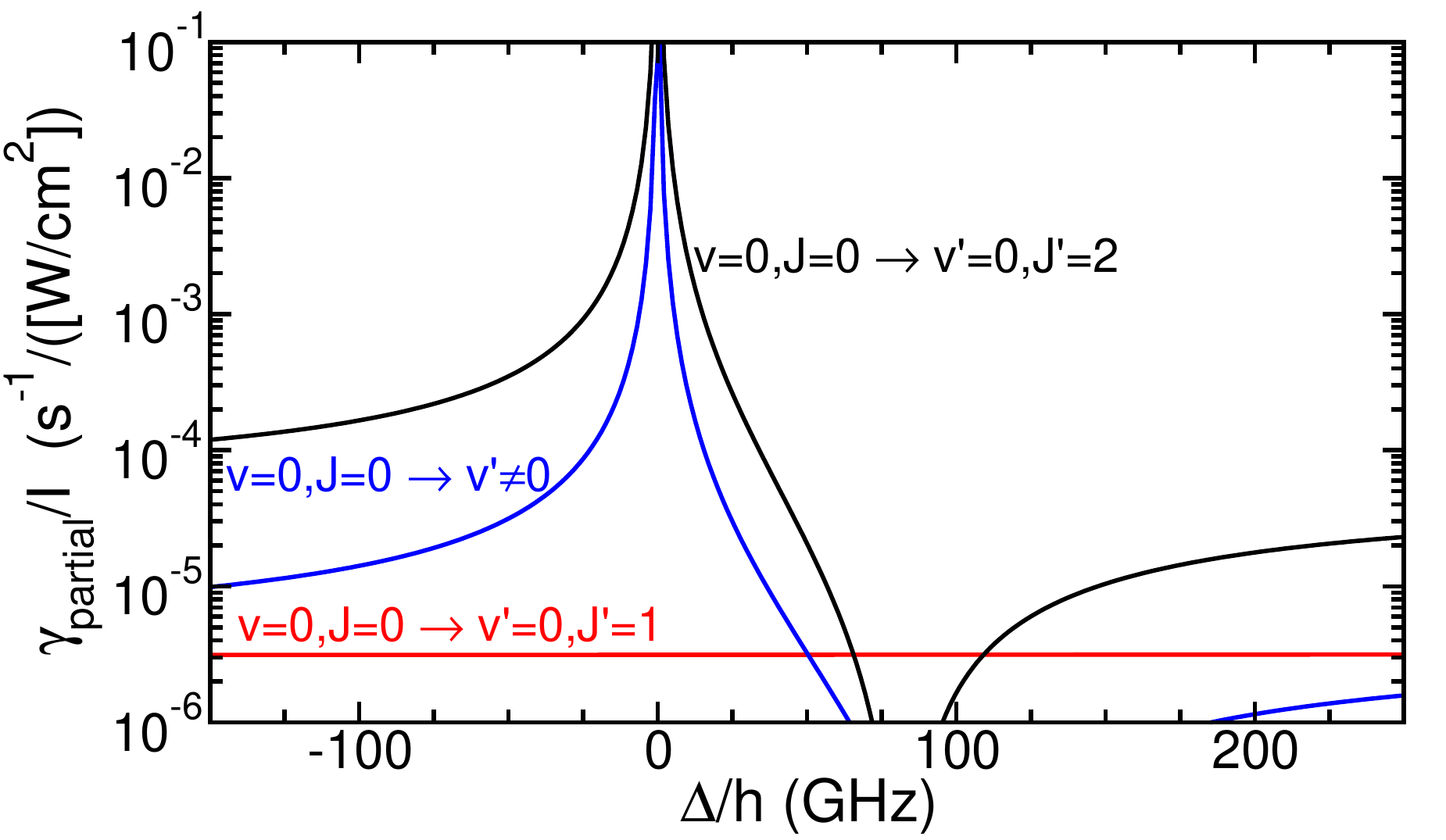}
\caption{Contributions to the Raman scattering rate divided by laser intensity, $\gamma_{\rm partial}/I$, of the ${v=0,J=0}$ rovibrational level of the X$^1\Sigma^+$ state of $^{23}$Na$^{87}$Rb
as functions of laser frequency detuning $\Delta/h$ from the  ${v''= 0,J''= 1}$ level of the coupled A$^1\Sigma^+$-b$^3\Pi_0$ system.
The red, black, and blue curves correspond to contributions to final rovibrational levels ${(v'=0,J'=1)}$, ${(v'=0,J'=2)}$,
and ${v'\ne 0}$ of the X$^1\Sigma^+$ state, respectively.
The laser is linearly polarized.  
}
\label{RamanJ0}
\end{figure}

In Fig.~\ref{RamanJ0} we plot three contributions to the Raman scattering rate 
summed over final state $M'$ projection, $\sum_{M'}\gamma_{{\rm X},v'J'M'\leftarrow {\rm X},vJM}$, for the ${v=0,J=0}$ (and thus $M=0$) rovibrational level of the X$^1\Sigma^+$ state of $^{23}$Na$^{87}$Rb as functions of laser frequency.
The sum of the three contributions corresponds to the total Raman scattering rate for this state.
Two of the three contributions have a resonant Fano line profile with a large value at the transition energy between the $v=0,J=0$ X$^1\Sigma^+$ level and the  ${v''= 0,J''= 1}$ level of the coupled A$^1\Sigma^+$-b$^3\Pi_0$ system
and a very small value to the blue of this transition.
The ${\rm X}, v=0,J=0\to {\rm X}, v'=0,J'=1$ contribution does not have a resonance and is constant over the range of frequencies shown in the figure.
Photon selections rules for this contribution imply that the A$^1\Sigma^+$-b$^3\Pi_0$ system does not contribute.
Finally, we observe that the sum of Raman rates to vibrational levels  $v'\ne 0$ of the X$^1\Sigma^+$ state is ten times 
smaller than the Raman rate to the $v'=0,J'=2$ rotational level of the X$^1\Sigma^+$ state (except where these rates are very small not shown in the figure.)

\begin{figure}
\includegraphics[scale=0.30,trim=0 0 0 0,clip]{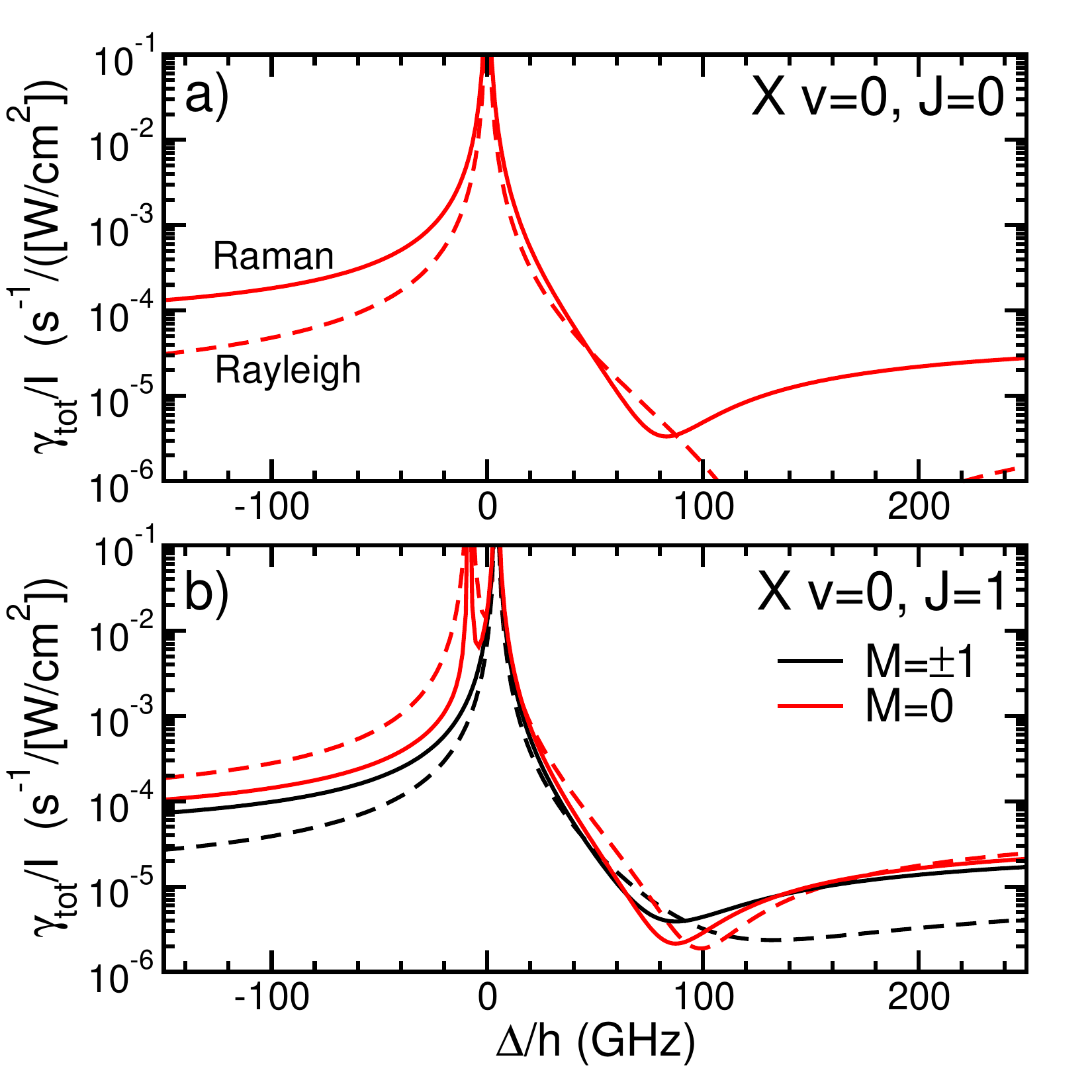}
\caption{Total Raman and Rayleigh scattering rates divided by laser intensity, $\gamma_{\rm tot}/I$, of the $v=0,J=0,M=0$ (panel a) and $v=0,J=1,M=\pm1,0$ (panel b) rovibrational levels of the X$^1\Sigma^+$ state of $^{23}$Na$^{87}$Rb
as functions of laser frequency detuning $\Delta/h$ from the $v=0,J=0$ X$^1\Sigma^+$ state to the  ${v''= 0,J''= 1}$ level of the coupled A$^1\Sigma^+$-b$^3\Pi_0$ system.
The solid and dashed curves in both panels correspond to Raman and Rayleigh scattering, respectively.
In panel b)  black and red curves correspond to rates for levels with projection quantum numbers $|M|=1$ and $M=0$,
respectively.
The laser is linearly polarized along the space-fixed quantization axis. }
\label{RamanRaleigh}
\end{figure}

We plot total Raman rate
\begin{equation}
 \gamma^{\rm Raman}_{{\rm X},vJM,{\rm tot}}  =\sum_{v'J'\ne vJ}\sum_{M'}\gamma_{{\rm X},v'J'M'\leftarrow {\rm X},vJM}
\end{equation}
and total Rayleigh rate
\begin{equation}
 \gamma^{{\rm Rayleigh}}_{{\rm X},vJM,{\rm tot}}  =\sum_{M'}\gamma_{{\rm X},vJM'\leftarrow {\rm X},vJM}
\end{equation}
for the $v=0,J=0$ and $v=0,J=1$ rovibrational levels of the X$^1\Sigma^+$ state of $^{23}$Na$^{87}$Rb in Fig.~\ref{RamanRaleigh}
as functions of laser frequency near the transition from the $v=0,J=0$ X$^1\Sigma^+$ state and the  ${v''= 0,J''= 1}$ level of the coupled A$^1\Sigma^+$-b$^3\Pi_0$ system  using a linear polarized laser photon along the $z$ axis. The Raman and Rayleigh rates are independent 
of the sign of projection quantum number $M$.
For the $v=0,J=0$ level the Raman rate is mostly smaller than the Raman rate, while for the  $v=0,J=1$ level these ratios depends on
projection quantum number $M$. The $v=0,J=1, M=\pm1$ rates are maximal for photon energy ${\cal E}_{v=0}+2B_{v=0}$, ignoring the small 
difference between the rotational constants of the $v=0$ level of the X$^1\Sigma^+$ state and the $v''=0$ level of the coupled A$^1\Sigma^+$-b$^3\Pi_0$ system. 
The ${v=0,J=1,M=0}$ rates are maximal for photon energies ${\cal E}_{v=0}-4B_{v=0}$ and ${\cal E}_{v=0}+2B_{v=0}$
again ignoring the small differences in rotational constants.
The two photon energies correspond to the $J''=0$ and 2 rotational states of $v''=0$ level of the coupled A$^1\Sigma^+$-b$^3\Pi_0$ system.
Crucially, the Raman and Rayleigh rates are smallest at a photon energy of $h\times 90$ GHz to the blue of the transition  between $v=0,J=0$ X$^1\Sigma^+$ state and the ${v''= 0,J''= 1}$ level of the coupled A$^1\Sigma^+$-b$^3\Pi_0$ system.
This detuning is close to the detuning for the magic conditions discussed in Fig.~\ref{magic}. In other words,
magic frequencies are also ``magic'' for decoherences.

\section*{Conclusion}

In this paper we have developed a theoretical approach to construct an optical trap for a single NaRb molecule where molecular rotational-hyperfine states have so called magic conditions. Constructing a rotational magic trap is the ideal solution the long rotational coherence times needed to exploit the rotational degree of
freedom as the quantum bit in quantum information processing. In such a laser trap, light-induced energy shifts of multiple rotational states of the ground configuration are the same, eliminating dephasing associated with spatial variations in intensity across the trap. This opens up the prospect of using the rotational degree of freedom of the molecule to encode a synthetic dimension in addition to having multiple molecule-containing traps in real space.

We reached this goal by changing the trapping laser frequency in a region that is close to or in between the narrow transitions from ${v = 0}, {J=0}$ of the X$^1\Sigma^+$ state to the ${v' = 0 }$ and ${v' = 1}$ vibrational levels of the spin-orbit coupled A$^1\Sigma^+$-b$^3\Pi_0$ complex. No external electric field is present while the magnetic field strength is 335.6 G. 
We predict nearly magic conditions for the lowest six rotational states of the $v=0$ level at  detuning $\Delta/h = -2$ GHz and 100 GHz from the $v'$=0, $J'=1$ level of the b$^3\Pi_0$ potential.
We have accounted for the nonzero nuclear spins of 
$^{23}$Na and $^{87}$Rb, which are aligned by the magnetic field through the Zeeman interaction. 
Finally, we have calculated  Raman and Rayleigh scattering rates in the off-resonant absorption of a laser photon by the alkali-metal molecule and we realized that these rates are smallest for laser frequencies where magic conditions hold near detunings of  100 GHz.

\section*{Acknowledgements}
Our research is supported by the U.S.~Air Force Office of Scientific Research Grants No.~FA9550-19-1-0272. Work at Temple University is also supported by the U.S. Air Force Office of Scientific Research Grants No.~FA9550-21-1-0153 and the NSF Grant No.~PHY-1908634.

\bibliography{DecoherencePolarizability}

\end{document}